\documentclass[sigconf,natbib=true, screen=true]{acmart}

\acmSubmissionID{2032}

\usepackage{hyperref}
\usepackage{amsmath}
\usepackage{graphicx}
\usepackage{multirow}
\usepackage{adjustbox}
\usepackage{tabularx}
\usepackage{booktabs}
\usepackage{tablefootnote}
\usepackage{booktabs}
\usepackage{array}
\usepackage{balance} 
\usepackage{multirow}
\usepackage[normalem]{ulem}
\usepackage{color}
\definecolor{lightgray}{RGB}{215,215,215}
\usepackage{colortbl}  
\usepackage{xcolor}
\useunder{\uline}{\ul}{}
\usepackage{subfigure}
\usepackage{algorithm}  
\usepackage{algorithmicx}  
\usepackage[noend]{algpseudocode}  

\usepackage{amsmath}  
\usepackage[inline]{enumitem}
\usepackage{tabularx}
\usepackage[utf8]{inputenc}
\usepackage[english]{babel}
\usepackage{amsthm}
\usepackage{bm}
\usepackage{acronym}
\usepackage{mdframed} 
\usepackage{lipsum}   
\usepackage[skip=2pt]{caption}
\newcommand{\mc}[1]{\mathcal{#1}}

\newcommand{\header}[1]{\vspace*{1mm}\noindent\textbf{#1.}}
\newtheorem{definition}{Definition}
\newtheorem{hypothesis}{Hypothesis}

\author{Zhen Zhang}
\orcid{0009-0001-0290-5386}
\affiliation{%
  \institution{Shandong University}
  \city{Qingdao}
  \country{China}
}
\email{zhen.zhang.sdu@gmail.com}

\author{Xinyu Ma}
\orcid{0000-0002-5511-9370}
\affiliation{%
  \institution{Baidu Inc.}
  \city{Beijing}
  \country{China}
}
\email{xinyuma2016@gmail.com}

\author{Weiwei Sun}
\orcid{0000-0002-4817-9500}
\affiliation{%
  \institution{Carnegie Mellon University}
  \city{Pittsburgh}
  \country{USA}
}
\email{sunnweiwei@gmail.com}

\author{Pengjie Ren}
\orcid{0000-0003-2964-6422}
\affiliation{%
  \institution{Shandong University}
  \city{Qingdao}
  \country{China}
}
\email{renpengjie@sdu.edu.cn}

\author{Zhumin Chen}
\orcid{0000-0003-4592-4074}
\affiliation{%
  \institution{Shandong University}
  \city{Qingdao}
  \country{China}
}
\email{chenzhumin@sdu.edu.cn}

\author{Shuaiqiang Wang}
\orcid{0000-0002-9212-1947}
\affiliation{%
  \institution{Baidu Inc.}
  \city{Beijing}
  \country{China}
}
\email{shqiang.wang@gmail.com}

\author{Dawei Yin}
\orcid{0000-0002-0684-6205}
\affiliation{%
  \institution{Baidu Inc.}
  \city{Beijing}
  \country{China}
}
\email{yindawei@acm.org}

\author{Maarten de Rijke}
\orcid{0000-0002-1086-0202}
\affiliation{%
  \institution{University of Amsterdam}
  \city{Amsterdam}
  \country{The Netherlands}
}
\email{m.derijke@uva.nl}

\author{Zhaochun Ren}
\authornote{Corresponding author.}
\orcid{0000-0002-9076-6565}
\affiliation{%
  \institution{Leiden University}
  \city{Leiden}
  \country{The Netherlands}
}
\email{z.ren@liacs.leidenuniv.nl}

\renewcommand{\shortauthors}{Zhen Zhang et al.}

\copyrightyear{2025}
\acmYear{2025}
\setcopyright{rightsretained}
\acmConference[SIGIR '25] {The 48th International ACM SIGIR Conference on Research and Development in Information Retrieval}{July 13--18, 2025}{Padua, Italy}
\acmISBN{}
\acmDOI{https://doi.org/10.1145/3726302.3730314}



\begin{document}

\title[Replication and Exploration of Generative Retrieval over Dynamic Corpora]{Replication and Exploration of\\ Generative Retrieval over Dynamic Corpora}

\begin{abstract}
Generative retrieval (GR) has emerged as a promising paradigm in information retrieval (IR).
However, most existing GR models are developed and evaluated using a static document collection, and their performance in dynamic corpora where document collections evolve continuously is rarely studied.
In this paper, we first reproduce and systematically evaluate various representative GR approaches over dynamic corpora.
Through extensive experiments, we reveal that existing GR models with \textit{text-based} docids show superior generalization to unseen documents.
We observe that the more fine-grained the docid design in the GR model, the better its performance over dynamic corpora, surpassing BM25 and even being comparable to dense retrieval methods.
While GR models with \textit{numeric-based} docids show high efficiency, their performance drops significantly over dynamic corpora.
Furthermore, our experiments find that the underperformance of numeric-based docids is partly due to their excessive tendency toward the initial document set, which likely results from overfitting on the training set.
We then conduct an in-depth analysis of the best-performing GR methods.
We identify three critical advantages of text-based docids in dynamic corpora:
(i) Semantic alignment with language models' pretrained knowledge,
(ii) Fine-grained docid design, and
(iii) High lexical diversity.
Building on these insights, we finally propose a novel multi-docid design that leverages both the efficiency of numeric-based docids and the effectiveness of text-based docids, achieving improved performance in dynamic corpus without requiring additional retraining. 
Our work offers empirical evidence for advancing GR methods over dynamic corpora and paves the way for developing more generalized yet efficient GR models in real-world search engines. Our code is available at \href{https://github.com/zhangzhen-research/dynamicGR}{here}.

\end{abstract}

\maketitle

\section{Introduction}
\label{sec:intro}
Traditional information retrieval (IR) systems are based on pipelined ``index-retrieval-then-rank'' strategies where each module is optimized separately~\cite{fan2022pre,lin2022pretrained,manning2009introduction}.
In recent years, a new retrieval paradigm, generative retrieval (GR) has emerged and garnered increasing attention due to its end-to-end training and remarkable performance~\cite{metzler2021rethinking, li2024matching, zhuang2022bridging, zhang2024generative}.
In this paradigm, all information of corpus is encoded into the parameters of a generative language model and such a model then directly generates document identifiers (docids) for a given query.
Specifically, generative retrieval primarily consists of two steps: 
(i) \textit{training as indexing}: through training objectives that map queries or documents to their associated docids, the model memorizes the document into its parameters during training and learns the mapping relationship between queries and docids.
(ii) \textit{generation as retrieval}: upon receiving a query, the model directly generates the corresponding docids autoregressively.
Existing work has shown that the design of docid plays a critical role in the performance of GR models~\cite{ziems2023large,zhou2023enhancing,tang2024generative}.
According to the type of docid, GR methods can be divided into two categories: \textit{numeric-based} methods that convert documents into numeric sequences via quantization strategies (hierarchical k-means~\cite{tay2022transformer}, product quantization~\cite{zhou2022ultron}, residual quantization~\cite{zeng2024scalable}), and \textit{text-based} methods that leverage titles, URLs, or n-grams as docids~\cite{de2020autoregressive,bevilacqua2022autoregressive}.

However, most GR approaches are built and evaluated in static scenarios, where the document collection remains fixed~\cite{jin2023language, yang2023auto, li2024distillation, tang2024listwise}. 
Although these models perform strongly in such settings, their effectiveness in real-world environments, where documents continuously evolve over time, has not been thoroughly evaluated~\cite{chen2023continual,kim2023exploring}. 
Unlike static corpora, dynamic corpora impose higher demands on a model's generalization and robustness as new documents are not encountered during training.
While previous studies have found that several GR models perform poorly over the dynamic corpora~\cite{kishore2023incdsi, chen2023continual}, only limited efforts have been made to comprehensively investigate the generalization ability of different GR models under fair experimental settings.

An intuitive solution is to retrain the GR model from scratch when the underlying corpus is updated.
However, the prohibitive computational costs of model training make this approach unfeasible~\cite{chen2022corpusbrain}.
Consequently, prior studies have explored continual (i.e., incremental) learning techniques to enhance the generalization ability of GR models with less computational overhead~\cite{mehta2022dsi++, guo2024corpusbrain++}. For example, \citet{mehta2022dsi++} proposes DSI++, which employs sharpness-aware minimization to optimize for flat loss basins, thereby enabling the model to effectively learn from newly added documents.
\citet{guo2024corpusbrain++} develop task-specific adapters~\cite{houlsby2019parameter,ma2022scattered} and pre-training objectives to adapt the dynamic corpora.
As shown in Table~\ref{tab:training}, we find that SEAL~\cite{bevilacqua2022autoregressive}, which employs text-based docids, can achieve comparable performance over dynamic corpora without re-training.

Therefore, our reproducibility study mainly focuses on the models' ability to generalize to unseen documents without additional training, and on analyzing the factors that influence the performance in dynamic corpora. 
Specifically, we first construct two dynamic corpora datasets based on the NQ~\cite{kwiatkowski2019natural} and MS-MARCO~\cite{bajaj2016ms} datasets, by partitioning the entire document collection and user queries into several sets, thereby simulating practical dynamic retrieval scenarios.
We then replicate various GR approaches and compare them with traditional retrieval models, including sparse retrieval and dense retrieval methods.
Our findings show that existing GR models exhibit inconsistent performance over dynamic scenarios. 
In comparison, GR models that use text-based docids outperform those that employ numeric-based docids in terms of effectiveness, and underperform them in terms of efficiency.

Building on the insights listed above, we analyze the advantages of text-based docids and limitations of numeric-based docids over dynamic corpora:
\vspace{-\topsep}
\begin{itemize}[leftmargin=*]
\item Models using numeric-based docids tend to generate docids that are encountered during initial training.
\item Text-based docids maintain semantic alignment with language models' pretraining distribution, enabling better generalization with unseen documents.
\item Text-based docids offer a finer-grained representation of documents, which enhances the model’s ability to generalize to unseen documents.
\item Text-based docids have higher lexical diversity, which helps mitigate overfitting to the initial training set.
\end{itemize}

\noindent%
Based on these observations, we design a novel multi-docid GR model that uses numeric-based docids but avoids the tendency to generate previously seen docids. It achieves finer granularity and larger docid size, and demonstrates superior performance and efficiency in dynamic corpora scenario.

In summary, our main contributions are as follows:
\begin{enumerate*}[label=(\roman*)]
    \item We introduce a comprehensive evaluation of existing GR approaches over dynamic corpora, focusing on their ability to generalize to unseen documents without additional training.
    \item We reveal the limitations of numeric-based docids in dynamic retrieval scenarios that they have a tendency toward the initial document set.
    \item We observed that the text-based docid performs better on dynamic corpora and analyzed key factors for its performance, including semantic alignment, fine-grained docids design, and high lexical diversity.
    \item Building on our findings, we propose a novel GR framework that integrates the high efficiency and low memory of numeric-based docids with the strong generalization capabilities of text-based docids.
\end{enumerate*}

\section{Background and Preliminaries}
\subsection{Generative retrieval}

The document retrieval task retrieves a relevant document \(d\) from a document collection \(\mc{D}\) given a user query \(q\). Traditional retrieval approaches typically rely on inverted indexing or similarity-based ranking; whereas GR retrieves documents by autoregressively generating the docid \(z\) corresponding to the most relevant document.

A GR model typically consists of two key components: an \textit{indexer} and a \textit{retriever}. 
The \textit{Indexer} acts as a document encoder that maps each document \(d \in \mc{D}\) to a unique identifier sequence (docid) \(\mathbf{z}=\{z_1, z_2, \dots, z_M\}\). 
Here each element \(z_t \in [K]\) is drawn a predefined vocabulary \(V\), which may include numerical tokens, lexical items, or other semantic identifiers. 
There are mainly two types of docid design: numeric-based and text-based:
\vspace{-\topsep}
\begin{itemize}[leftmargin=*]
\item numeric-based: Each document corresponds to a numeric sequence, typically generated by converting document embeddings into numeric sequences using clustering or quantization methods.
\item text-based: Uses metadata from documents as docid, generally including elements such as title, query, URL, n-gram, etc.
\end{itemize}

\noindent%
As for the \textit{retriever}, upon taking a user query \(q \in \mathcal{Q}\), the \textit{Retriever} generates the most relevant docid by maximizing the conditional probability over the sequence of tokens representing the docid:
\begin{equation}
P(z_1, z_2, \dots, z_M \mid q; \theta) = \prod_{t=1}^{M} P\bigl(z_t \mid z_1, \dots, z_{t-1}, q; \theta\bigr),
\end{equation}
where \(\theta\) denotes the model parameters, \(M\) is the predefined docid length, and \(\{z_1, \dots, z_{t-1}\}\) represents the previously generated tokens. 
During inference, the system performs beam search to find:
\begin{equation}
\hat{\mathbf{z}} = \mathop{\text{argmax}}_{\mathbf{z} \in \mc{T}_\mc{D}} P(z_{1:M}|q;\theta),
\end{equation}
where a prefix tree (\(\mc{T}_\mc{D}\)) enforces constrained decoding by storing all valid docid prefixes from the corpora \(\mc{D}\). At each decoding step \(t\), token selection is restricted to the children nodes of the current prefix \(z_{1:t-1}\) in \(\mc{T}_\mc{D}\), ensuring generated sequences correspond to existing documents.

\subsection{Task formulation}

The dynamic corpora involves continuously adapting to an expanding document collection while maintaining retrieval capabilities.
Initially, we are given:
\begin{itemize}[leftmargin=*]
    \item an initial document set $\mc{D}_0 = \{d_1, d_2, \dots, d_n\}$, and
    \item corresponding query-document pairs $\mc{P}_0 = \{ \langle q_i, d_j \rangle \mid q_i \in \mc{Q}, d_j \in \mc{D}_0 \}$,
\end{itemize}
where each document $d_j \in \mc{D}_0$ is a text sequence and $\mc{Q} = \{q_1, q_2$, \ldots, $q_m\}$ denotes the initial query set. A GR model is first trained using standard sequence-to-sequence objectives on $\mc{P}_0$.
The core challenge arises with incremental updates: when new documents $\mc{D}_{\text{new}} = \{d_{n+1}, \dots, d_{n+k}\}$ are introduced.
For each new document $d_{n+i} \in \mc{D}_{\text{new}}$, the \textit{indexer} generates a corresponding docid:
\begin{equation}
\mathbf{z}_{n+i} = \{z^{(n+i)}_1, \dots, z^{(n+i)}_M\},\quad z^{(n+i)}_t \in \{1,\dots,K\},
\end{equation}
preserving the original token vocabulary and sequence length $M$. These new identifiers are incorporated into the existing prefix tree $\mc{T}_\mc{D}$.

The \textit{retriever} must then handle the expanded search space $\mc{D}' = \mc{D}_0 \cup \mc{D}_{\text{new}}$. 
Two adaptation strategies emerge:
\begin{enumerate*}[label=(\roman*)]
    \item \textbf{Model adaptation}: Continuously train or retrain the model on $\mc{P}_0 \cup \mathcal{P}_{\text{new}}$ to learn updated document representations.
    \item \textbf{Index adaptation}: Maintain frozen model parameters while updating only the indexing structures.
\end{enumerate*}
As mentioned in the Introduction Section, retraining or continuing training the GR model would cost much more computational overhead.
Therefore, our work only focuses on the second paradigm, where the \textit{Retriever} must leverage learned knowledge to handle novel documents through constrained decoding over the updated prefix tree $\mc{T}_\mc{D}'$.
This setting tests the model's ability to generalize to unseen document representations without parameter updates.

\subsection{Evaluation metrics}
Our evaluation primarily focuses on two aspects of model performance over dynamic corpora:  
\begin{enumerate*}[label=(\roman*)]
\item \textit{retrieve initial documents}, which assesses the model's ability to maintain performance on queries targeting original documents \( \mc{D}_0 \); and  
\item \textit{retrieve newly added documents}, which evaluates the model's capacity to retrieve novel documents \( \mc{D}_{\text{new}} \).  
\end{enumerate*}
We use \texttt{Hit@10} as the primary metric in both settings and further introduce formal metrics to summarize the model’s performance as new documents are incrementally indexed.

\header{Retrieve initial documents}  
To assess the model's forgetting behavior when indexing new documents, we define the forgetting metric \( F_n \), which quantifies the degradation in retrieval performance on queries targeting the original corpus \( \mc{D}_0 \) after indexing corpus \( \mc{D}_1 \) to \( \mc{D}_n \):
\begin{definition}[Forgetting Metric \( F_n \)]
\begin{equation}
F_n = \frac{1}{n} \sum_{o=1}^{n} \max\left( P_{0,0} - P_{o,0}, \, 0 \right)
\end{equation}
where \( P_{o,0} \) represents the retrieval performance (e.g., Hit@10) of queries in \( \mc{D}_0 \) after indexing corpus \( \mc{D}_o \).
\end{definition}

\header{Retrieve newly added documents}  
To measure the model’s ability to generalize and retrieve newly indexed documents, we define the generalization performance metric \( GA_n \), which captures how well the model retrieves queries associated with incrementally added documents:
\begin{definition}[Generalization Performance \( GA_n \)]
\begin{equation}
GA_n = \frac{1}{n} \sum_{o=1}^{n} P_{o,o}
\end{equation}
where \( P_{o,o} \) represents the retrieval performance on queries targeting \( \mc{D}_o \) after indexing corpus \( \mc{D}_o \).
\end{definition}

\section{Experimental Setup}
\label{sec:exper}

\subsection{Datasets}
We conduct our experiments on two widely used datasets: Natural Questions (NQ)~\cite{kwiatkowski2019natural} and MS-MARCO~\cite{bajaj2016ms}. 
We appropriately partition the document sets within the datasets to simulate dynamic corpora scenarios.


To simulate the task of dynamic corpora, we partition the documents sets in MS-MARCO and NQ through a two-phase process: 
\begin{enumerate*}[label=(\roman*)]
\item \textbf{Initial corpus construction:}
   Collect all documents and randomly select 50\% as the initial corpus \( \mc{D}_0 \).
   Extract query-document pairs associated with \( \mc{D}_0 \) and split them into a training set \( \mc{P}_0 \) and an initial test set \( \mc{Q}_0 \). All models are fully trained on \( \mc{P}_0 \), simulating initialization on the base corpus.
\item \textbf{Incremental document additions:}
   Partition the remaining 50\% of documents into five equally sized subsets, each comprising 10\% of the original corpus (\( \mc{D}_1 \) to \( \mc{D}_5 \)).
   For each \( \mc{D}_i \) (\( 1 \leq i \leq 5 \)), extract its associated query-document pairs to form incremental test sets \( \mc{Q}_1 \) to \( \mc{Q}_5 \). No additional training is performed for these document sets—models only index new documents upon each addition.
\end{enumerate*}

\subsection{Retrieval approaches}
To compare the performance of GR approaches with previous retrieval methods on dynamic corpora, we selected three types of retrieval approaches: sparse retrieval approaches, dense retrieval approaches, and generative retrieval approaches.

\header{Sparse retrieval}
\begin{enumerate*}[label=(\roman*)] 
    \item BM25~\cite{robertson2009probabilistic}, a traditional retrieval approach that ranks documents according to the frequency of the term and the normalization of the length of the document. We re-index the entire corpora when the corpora is updated.
\end{enumerate*}

\begin{table*}[h]
\centering
\caption{Model performance on queries corresponding to the initial document set $\mc{D}_0$, where the document set size continuously expands as new documents are added. The best results are indicated in boldface.}
\setlength{\tabcolsep}{1.2mm}
\begin{tabular}{@{}l l ccccccc ccccccc@{}}
\toprule
\multirow{2}{*}{Method} & \multirow{2}{*}{DocID Type} & \multicolumn{7}{c}{NQ (Hit@10)} & \multicolumn{7}{c}{MS-MARCO (Hit@10)} \\ 
\cmidrule(lr){3-9} \cmidrule(lr){10-16}
 & & $\mc{D}_0$ & $\mc{D}_1$ & $\mc{D}_2$ & $\mc{D}_3$ & $\mc{D}_4$ & $\mc{D}_5$ & $F_n$ $\downarrow$ & $\mc{D}_0$ & $\mc{D}_1$ & $\mc{D}_2$ & $\mc{D}_3$ & $\mc{D}_4$ & $\mc{D}_5$  & $F_n$ $\downarrow$ \\ \midrule
\multicolumn{16}{@{}l}{\textbf{Sparse retrieval}} \\ 
BM25 & Term Weight & 0.647 & 0.625 & 0.611 & 0.598 & 0.573 & 0.573 & 0.051 & 0.653 & 0.640 & 0.632 & 0.629 & 0.619 & 0.614 & 0.026 \\ \midrule
\multicolumn{16}{@{}l}{\textbf{Dense retrieval}} \\ 
DPR & Dense Vector & 0.725 & 0.704 & 0.696 & 0.686 & 0.670 & 0.660 & 0.042 & 0.683 & 0.681 & 0.668 & 0.656 & 0.651 & 0.648 & 0.022 \\ 
DPR-HN & Dense Vector & 0.826 & 0.801 & 0.797 & 0.776 & 0.773 & 0.768 & 0.043 & \textbf{0.723} & \textbf{0.712} & \textbf{0.692} & \textbf{0.685} & \textbf{0.672} & \textbf{0.664} & 0.038 \\ \midrule
\multicolumn{16}{@{}l}{\textbf{Generative retrieval}} \\ 
DSI-SE & Category Nums & 0.718 & 0.710 & 0.706 & 0.702 & 0.699 & 0.696 & \textbf{0.015} & 0.605 & 0.601 & 0.597 & 0.594 & 0.592 & 0.589 & \textbf{0.010} \\ 
Ultron-PQ & Category Nums & 0.795 & 0.785 & 0.780 & 0.780 & 0.762 & 0.755 & 0.023 & 0.663 & 0.655 & 0.647 & 0.643 & 0.637 & 0.632 & 0.020 \\ 
NCI & Category Nums & \textbf{0.871} & 0.856 & 0.844 & \textbf{0.839} & 0.811 & 0.802 & 0.041 & 0.702 & 0.693 & 0.673 & 0.667 & 0.654 & 0.633 & 0.038 \\ 
GenRET & Category Nums & 0.858 & 0.853 & 0.836 & 0.829 & 0.812 & 0.796 & 0.033 & 0.717 & 0.697 & 0.688 & 0.674 & 0.659 & 0.652 & 0.043 \\ 
Ultron-URL & URL Path & 0.816 & 0.810 & 0.794 & 0.781 & 0.780 & 0.768 & 0.029 & 0.626 & 0.620 & 0.618 & 0.614 & 0.611 & 0.608 & 0.012 \\ 
SEAL & N-gram & 0.809 & 0.806 & 0.788 & 0.774 & 0.774 & 0.763 & 0.028 & 0.661 & 0.641 & 0.625 & 0.616 & 0.602 & 0.598 & 0.045 \\ 
MINDER & Multi-text & 0.838 & 0.828 & 0.813 & 0.811 & 0.801 & 0.773 & 0.033 & 0.667 & 0.649 & 0.633 & 0.625 & 0.612 & 0.600 & 0.043 \\ 
LTRGR & Multi-text & 0.862 & \textbf{0.857} & \textbf{0.846} & 0.827 & \textbf{0.813} & \textbf{0.807} & 0.032 & 0.688 & 0.675 & 0.660 & 0.649 & 0.636 & 0.621 & 0.040 \\ 
\bottomrule
\end{tabular}
\label{tab:comparison1}
\end{table*}

\header{Dense retrieval}
\begin{enumerate*}[label=(\roman*)] 
    \item DPR~\cite{karpukhin2020dense}, a neural retrieval method that uses dual encoders to map queries and documents into a shared dense vector space for similarity computation. We use the document encoder trained on \( \mc{D}_0 \) to re-encode the newly added documents for retrieval.
    \item DPR-HN~\cite{qu2020rocketqa,ma2022pre}, an enhanced DPR variant that incorporates hard negative sampling techniques. It integrates in-batch negatives, top-K retrieved negatives from dense retrievers, and BM25 hard negatives.
\end{enumerate*}

\header{Generative retrieval}
\begin{enumerate*}[label=(\roman*)] 
    \item DSI-SE~\cite{tay2022transformer}, which uses the hierarchical k-means clustering results of document embeddings as docids and trains the model to memorize the documents in parameters.
    \item Ultron-PQ~\cite{zhou2022ultron} uses the product quantization results of document embeddings as docids and designs a three-stage training task to enable the model to memorize the documents.
    \item Ultron-URL~\cite{zhou2022ultron}, a variant of Ultron, uses the document URLs as docids.
    \item NCI~\cite{wang2022neural} uses a prefix-aware weight-adaptive decoder and various query generation strategies to train the model.
    \item GenRET~\cite{sun2024learning} uses constrained cluster centroids as docids and trains the representations of docids through document tokenization and document reconstruction tasks.
    \item SEAL~\cite{bevilacqua2022autoregressive} uses arbitrary n-grams from documents as docids and generates a series of n-grams under the constraints of the FM-index to retrieve the corresponding documents.
    \item MINDER~\cite{li2023multiview} adopts multiple text types to represent docids, such as titles, URLs, and n-grams. Different types of scores are generated at the same time and documents are retrieved based on these scores.
    \item LTRGR~\cite{li2024learning} also adopts the multiple text docid design, and introduces an additional learn-to-rank task and rank loss to optimize the retrieval model.
\end{enumerate*}

To enable GR approaches to handle dynamic corpora, we adopt the following design strategies for docid assignment: 
\begin{enumerate*}[label=(\roman*)] 
\item\textbf{For numeric-based docids}: These approaches rely on structured numerical representations (e.g., k-means centroids, product quantization codebooks). 
We preserve the original document encoder's state (k-means cluster centroids or vector quantization codebooks) to encode new documents, maintaining identifier consistency with initial documents.  
\item\textbf{For text-based docids}: 
These approaches leverage document-induced textual patterns (e.g., Title, URLs).
We leverage new documents' inherent metadata and text structure to automatically assemble valid identifiers. 
\end{enumerate*}

\subsection{Implementation details}
We implement BM25 using the \texttt{bm25s} library.\footnote{\url{https://github.com/xhluca/bm25s}} For dense retrieval models (DPR and DPR-HN), we employ the \texttt{pyserini} toolkit~\cite{lin2021pyserini}, using its built-in functionalities for indexing and retrieval. For GR approaches (DSI, ULtron variants, NCI, GenRET, SEAL, MINDER, and LTRGR), we adopt their official implementations and strictly follow the default hyperparameter configurations provided in the original works to ensure reproducibility.  

All models operate with a maximum input sequence length of 512 tokens. Experiments are conducted on 8 NVIDIA A100 GPUs, with distributed training enabled for GR methods to accommodate their large parameter sizes. For DPR/DPR-HN, we initialize the document encoder with the checkpoint pre-trained on the initial documents and freeze its parameters during incremental phases.

\subsection{Statistical Validation}
We verified the reliability of retrieval performance.
All experimental results reported in Tables~\ref{tab:comparison1} and Table~\ref{tab:comparison2} achieved statistical significance at \( p < 0.05 \).

\section{Performance over Dynamic Corpora}
\begin{table*}[h]
\centering
\caption{Model performance on queries corresponding to the newly added document set $\mc{D}_1$ to $\mc{D}_5$, where the document set size continuously expands as new documents are added. The best results are indicated in boldface.}
\setlength{\tabcolsep}{1.2mm}
\begin{tabular}{@{}l l ccccccc ccccccc@{}}
\toprule
\multirow{2}{*}{Method} & \multirow{2}{*}{DocID Type} & \multicolumn{7}{c}{NQ (Hit@10)} & \multicolumn{7}{c}{MS-MARCO (Hit@10)} \\ 
\cmidrule(lr){3-9} \cmidrule(lr){10-16}
 & & $\mc{D}_0$ & $\mc{D}_1$ & $\mc{D}_2$ & $\mc{D}_3$ & $\mc{D}_4$ & $\mc{D}_5$ & $GA_n$ $\uparrow$ & $\mc{D}_0$ & $\mc{D}_1$ & $\mc{D}_2$ & $\mc{D}_3$ & $\mc{D}_4$ & $\mc{D}_5$  & $GA_n$ $\uparrow$ \\ \midrule
\multicolumn{16}{@{}l}{\textbf{Sparse retrieval}} \\ 
BM25 & Term Weight & 0.647 & 0.620 & 0.588 & 0.598 & 0.552 & 0.571 & {0.586} & 0.653 & 0.634 & 0.631 & 0.620 & 0.603 & 0.601 & {0.618} \\ \midrule

\multicolumn{16}{@{}l}{\textbf{Dense retrieval}} \\ 
DPR & Dense Vector & 0.725 & 0.580 & 0.587 & 0.570 & 0.531 & 0.544 & {0.562} & 0.683 & 0.625 & 0.623 & 0.599 & 0.607 & 0.604 & {0.612} \\ 
DPR-HN & Dense Vector & 0.826 & 0.645 & 0.644 & 0.626 & 0.621 & 0.624 & 0.632 & \textbf{0.723} & \textbf{0.662} & \textbf{0.653} & \textbf{0.642} & \textbf{0.623} & \textbf{0.619} & {\textbf{0.640}} \\ \midrule

\multicolumn{16}{@{}l}{\textbf{Generative retrieval}} \\ 
DSI-SE & Category Nums & 0.718 & 0.231 & 0.203 & 0.221 & 0.185 & 0.205 & {0.209} & 0.605 & 0.204 & 0.197 & 0.186 & 0.172 & 0.159 & {0.184} \\ 
Ultron-PQ & Category Nums & 0.795 & 0.548 & 0.549 & 0.542 & 0.539 & 0.532 & {0.542} & 0.663 & 0.428 & 0.415 & 0.399 & 0.384 & 0.376 & {0.400} \\ 
NCI & Category Nums & \textbf{0.871} & 0.464 & 0.437 & 0.433 & 0.358 & 0.323 & {0.403} & 0.702 & 0.402 & 0.380 & 0.355 & 0.341 & 0.320 & {0.360} \\ 
GenRET & Category Nums & 0.858 & 0.361 & 0.419 & 0.401 & 0.357 & 0.354 & {0.378} & 0.717 & 0.439 & 0.425 & 0.396 & 0.350 & 0.331 & {0.388} \\ 
Ultron-URL & URL Path & 0.816 & 0.553 & 0.545 & 0.543 & 0.541 & 0.532 & {0.543} & 0.626 & 0.397 & 0.376 & 0.364 & 0.354 & 0.342 & {0.367} \\ 
SEAL & N-gram & 0.809 & 0.744 & 0.736 & 0.727 & 0.727 & 0.725 & {0.732} & 0.661 & 0.611 & 0.607 & 0.584 & 0.571 & 0.559 & {0.586} \\ 
MINDER & Multi-text & 0.838 & 0.803 & 0.751 & 0.746 & 0.742 & 0.736 & {0.756} & 0.667 & 0.614 & 0.608 & 0.587 & 0.569 & 0.546 & {0.585} \\ 
LTRGR & Multi-text & 0.862 & \textbf{0.831} & \textbf{0.803} & \textbf{0.811} & \textbf{0.779} & \textbf{0.773} & {\textbf{0.799}} & 0.688 & 0.621 & 0.612 & 0.601 & 0.589 & 0.577 & {0.600} \\
\bottomrule
\end{tabular}
\label{tab:comparison2}
\end{table*}

In this section, we analyze the experimental results in dynamic corpora scenario, focusing on how different retrieval approaches perform as the document set expands incrementally. 

\header{Retrieving initial documents}
We investigate how the incremental indexing of new documents affects the retrieval performance of queries corresponding to the initial document set \( \mc{D}_0 \), as the document collection grows.
Both BM25 and DPR exhibit stable performance in maintaining retrieval effectiveness for initial documents \( \mc{D}_0 \).

Generative retrieval methods demonstrate better resistance to forgetting.
For example, DSI-SE achieves \( F_n \) values (0.015 on NQ and 0.010 on MS-MARCO), outperforming dense retrieval approaches.
Because there is no additional training step, the mapping from documents to docids remains intact, enabling these models to maintain consistent performance for the initially indexed documents.
However, a potential drawback lies in their limited flexibility when new documents are introduced, as these methods often struggle to index unseen docids effectively unless specific adaptations (e.g., additional training) are employed.

\header{Retrieving newly added documents}
When retrieving newly added documents from a dynamically expanding document collection, 
both BM25 and DPR demonstrate stable generalization performance (\(GA_n\)) when retrieving newly added documents, as measured by their ability to adapt to an expanding corpus. BM25 achieves \(GA_n\) scores of 0.586 on NQ and 0.618 on MS-MARCO, while DPR attains 0.562 and 0.612 respectively.

Generative retrieval models exhibit huge divergence in generalization performance (\(GA_n\)): 
Numeric-based docid (e.g., DSI-SE, Ultron-PQ) demonstrate critical limitations in adapting to unseen documents.
DSI-SE’s \(GA_n\) values (0.209 on NQ, 0.184 on MS-MARCO) align with its severe Hits@10 degradation when retrieving new documents (e.g., dropping from 0.718 to 0.205 on NQ’s \( \mc{D}_5 \)). This failure stems from rigid numeric docid mappings learned during training, which lack inherent semantic connections to new content. Similarly, Ultron-PQ’s numeric identifiers yield unstable \(GA_n\) (0.542 on NQ, 0.400 on MS-MARCO), as its quantization-based docid system struggles to encode novel document semantics without retraining. 

Text-based docid methods, such as SEAL and LTRGR, demonstrate much more promising results. 
For example, LTRGR attain \(GA_n = 0.799\) on NQ, with only a 0.089 Hits@10 drop from \( \mc{D}_0 \) to \( \mc{D}_5 \), outperforming even optimized dense retrieval (DPR-HN: \(GA_n = 0.632\)). 
SEAL’s n-gram docids similarly excel (\(GA_n = 0.732\) on NQ).
This suggests that text-based docids, such as n-grams or titles, offer more flexibility and generalization.
Text-based docids can inherently adapt to newly added documents, which often contain similar linguistic features. Consequently, models using text-based docids are better equipped to maintain high retrieval performance as the corpus evolves.
These models can leverage the semantic richness of text-based representations, allowing them to effectively handle the continuous expansion of the document set without significant performance loss.

\header{Incremental training vs direct generalization}
\begin{table}[t]
\centering
\setlength{\tabcolsep}{6pt}
\caption{Comparison of retrieval performance (Hit@10) on the NQ dataset for initial and newly added documents.}
\begin{tabular}{@{}lcccccc@{}}
\toprule
\multirow{2}{*}{\textbf{Model}} & \multicolumn{6}{c}{\textbf{NQ (Hit@10)}} \\
\cmidrule(lr){2-7}
& $\mc{D}_0$ & $\mc{D}_1$ & $\mc{D}_2$ & $\mc{D}_3$ & $\mc{D}_4$ & $\mc{D}_5$ \\
\midrule
\multicolumn{7}{@{}l}{\emph{Initial documents}} \\
\midrule
DSI    & 0.718 & 0.710 & 0.706 & 0.702 & 0.699 & 0.696 \\
DSI++  & 0.718 & 0.697 & 0.687 & 0.682 & 0.676 & 0.673 \\
SEAL   & 0.809 & \textbf{0.806} & 0.788 & 0.774 & 0.774 & 0.763 \\
SEAL++ & \textbf{0.809} & 0.791 & \textbf{0.788} & \textbf{0.781} & \textbf{0.776} & \textbf{0.766} \\
\midrule
\multicolumn{7}{@{}l}{\emph{Newly added documents}} \\
\midrule
DSI    & 0.718 & 0.231 & 0.203 & 0.221 & 0.185 & 0.205 \\
DSI++  & 0.718 & 0.677 & 0.671 & 0.667 & 0.657 & 0.644 \\
SEAL   & 0.809 & 0.744 & 0.736 & 0.727 & 0.727 & 0.725 \\
SEAL++ & \textbf{0.809} & \textbf{0.768} & \textbf{0.756} & \textbf{0.744} & \textbf{0.743} & \textbf{0.733} \\
\bottomrule
\end{tabular}
\label{tab:training}
\end{table}

We implemented incremental training inspired by the DSI++ framework for both DSI and SEAL. Our approach involves two key components: (i) incremental indexing training, where models are fine-tuned on newly added documents to learn their docid mappings, and (ii) random replay training, which reintroduces a subset of historical documents during fine-tuning to mitigate forgetting. As shown in Table~\ref{tab:training}, incremental training significantly enhances the models' ability to retrieve newly added documents for DSI. For instance, DSI++ improves Hits@10 for new documents on NQ from 0.205 to 0.644 on $ \mc{D}_5 $, demonstrating better generalization to unseen documents. However, this comes at a cost: DSI++ exhibits noticeable forgetting on initial documents.
For SEAL, incremental training yields only a slight improvement in both settings.
This is because SEAL's text-based docids inherently accommodate semantic variations in new documents, reducing the need for extensive retraining. 
Further analysis in Table~\ref{tab:training} suggests that while incremental training provides modest benefits for text-based docid models, their robustness to dynamic corpora primarily stems from their ability to generalize via natural language patterns, rather than relying on explicit retraining.

In conclusion, the performance of generative retrieval models in dynamic corpora reveals certain limitations, particularly when it comes to adapting to newly added documents.
For retrieving initial documents, generative models perform acceptably, with only slight degradation as the corpus grows. Models like DSI-SE and NCI show only slight performance drops, similar to traditional methods like BM25 and DPR.
However, the challenge arises when retrieving newly added documents. Generative models that rely on numeric-based docids (e.g., DSI-SE and Ultron-PQ) struggle significantly, as they cannot adapt to new documents without retraining. In contrast, models using text-based docids, like SEAL and LTRGR, perform better, as text-based docids can generalize and adapt to new documents more effectively.
While text-based docids show promise, not all text-based generative retrieval models perform equally well. For instance, models like SEAL demonstrate better performance compared to others like Ultron-URL. Further exploration is needed to determine which specific text features are most suitable for dynamic corpora scenario.

\section{Analysis of Docid Design}
\label{sec:factors}

To understand the pros and cons of different docid designs for GR in dynamic corpus scenarios, we conduct a series of experiments.
In Section \ref{sec:bias}, we introduce the Initial Document Bias Index (IDBI) to analyze the bias of different GR methods towards older documents when new documents are added. We observe that numeric-based docids exhibit significantly higher bias compared to text-based docids, which, to some extent, explains the poor performance of these methods in dynamic corpus settings.
Then, in Section~\ref{sec:docid-type}, we conduct a comprehensive ablation study on text-based docid methods to investigate how docid type, granularity, and vocabulary size affect models’ generalization to new documents. Our analysis shows that more semantic, finer-grained, and larger vocabulary choices often lead to superior results.

\subsection{Bias to initial documents}
\label{sec:bias}

To explain the poor performance of numeric-based docids on new documents, we hypothesize as follows:

\begin{hypothesis}[Semantic Familiarity]
The effectiveness of generative retrieval (GR) on new documents is correlated with 
how well the docid representations align with the language model’s pretraining distribution. Formally, let \( P_{\text{LM}}(x) \) denote the token distribution learned by the language model over the vocabulary \( \mathcal{V} \), and let \( P_{\text{docid}}(x) \) represent the probability distribution induced by the docid representation space. We define the semantic familiarity of a docid system as:
\begin{equation}
    \mathcal{S} = \mathbb{E}_{x \sim P_{\text{docid}}} \left[ \log P_{\text{LM}}(x) \right].
\end{equation}
A higher value of \( \mathcal{S} \) indicates better alignment between the docid distribution and the language model’s pretraining distribution.
\end{hypothesis}


\begin{figure}[t]
  \centering
  \begin{tabular}{c}
  \includegraphics[clip, trim=70mm 165mm 70mm 0mm, width=\linewidth]{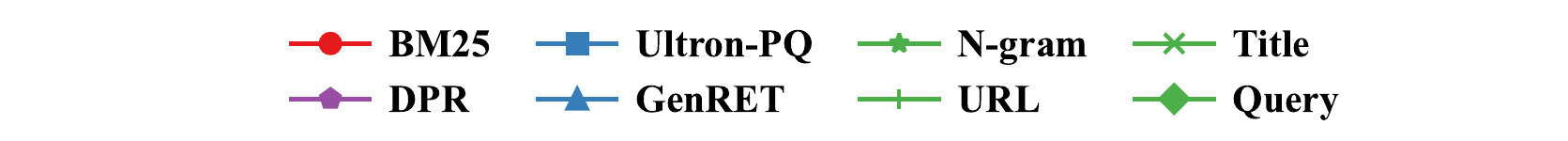}
  \\
  \includegraphics[clip, trim=0mm 0mm 220mm 22mm,width=0.9\linewidth, height=0.6\linewidth]{Figures/06-proportion.pdf}
  \\
  \includegraphics[clip, trim=220mm 0mm 0mm 22mm,width=0.9\linewidth, height=0.6\linewidth]{Figures/06-proportion.pdf}
  \end{tabular}
  \caption{IDBI results for different retrieval methods on NQ dataset. Lower is better.}
  \label{fig:proportion}
\end{figure}

Text-based docids inherently preserve distributional alignment with the underlying language model. By leveraging substrings (e.g., n-grams) extracted from document text, these identifiers ensure lexical overlap between new documents and the model’s pretraining data, thereby maintaining high semantic familiarity. 

In contrast, numeric-based docids, despite capturing latent semantic structures through clustering or vector quantization, form a distinct symbolic system that lacks grounding in the model’s pretraining data. This misalignment impairs the model's ability to associate a new document with its identifier, particularly in scenarios where the document itself has never been encountered before. As a result, generative retrieval systems exhibit retrieval bias, favoring initial documents over newly introduced ones.

To quantify the retrieval bias toward initial documents, we introduce the \textbf{Initial Document Bias Index (IDBI)}.

\begin{definition}[Initial Document Bias Index (IDBI)]
The \textbf{IDBI} measures the discrepancy between the observed proportion of initial documents retrieved in the Top-$K$ results and their expected proportion under an unbiased distribution. It is defined as:
\begin{equation}
\text{IDBI} = \frac{R_{\text{init}} - E_{\text{init}}}{K - E_{\text{init}}},
\end{equation}
where: \( R_{\text{init}} \) is the number of retrieved initial documents in the Top-$K$ results. \( E_{\text{init}} = K \times \frac{|\mathcal{D}_0|}{|\mathcal{D}_0 \cup \mathcal{D}_{\text{new}}|} \) is the expected count of initial documents under a uniform distribution.
\end{definition}
\textit{Remark.}
The IDBI assumes that initial documents \( \mathcal{D}_0 \) and new documents \( \mathcal{D}_{\text{new}} \) are drawn from the same latent query-document relevance distribution. The index is normalized within the range \([0,1]\), where:
\( \text{IDBI} = 0 \) indicates no bias (i.e., retrieved documents are proportional to their corpus distribution).
\( \text{IDBI} = 1 \) indicates complete bias toward initial documents.

As shown in Figure~\ref{fig:proportion}, numeric-based docids (e.g., GenRET, Ultron-PQ) exhibit significant retrieval bias toward initial documents across both task settings. This aligns with our hypothesis that numeric-based representations introduce a semantic gap, making it harder for the model to associate new documents with their identifiers.
While text-based docids (e.g., URL, title, query) demonstrate better semantic alignment, they still exhibit retrieval bias, particularly for titles and URLs. This is likely due to their abstract nature, which fails to capture fine-grained document content.
In contrast, BM25 and DPR, which perform direct text similarity matching without relying on predefined docids, show minimal retrieval bias. Their ability to retrieve new documents equitably stems from their reliance on content-based representations rather than fixed identifier mappings.
N-gram docids achieve comparable performance by structurally aligning with the generative model’s learning paradigm. Unlike titles or URLs, which require abstract semantic interpretation, n-grams preserve raw token sequences, which the model inherently optimizes for during generation. This alignment ensures that even previously unseen documents benefit from the model’s pre-existing familiarity with local linguistic patterns.

\begin{table}[t]
\centering
\setlength{\tabcolsep}{6pt} 
\caption{Comparison of retrieval performance (Hit@10) on the NQ dataset for initial and newly added documents across various text-type docid representations.}
\begin{tabular}{@{}lcccccc@{}}
\toprule
\multirow{2}{*}{\textbf{Text Type}} & \multicolumn{6}{c}{\textbf{NQ (Hit@10)}} \\
\cmidrule(lr){2-7}
& $\mc{D}_0$ & $\mc{D}_1$ & $\mc{D}_2$ & $\mc{D}_3$ & $\mc{D}_4$ & $\mc{D}_5$ \\
\midrule
\multicolumn{7}{@{}l}{\emph{Initial documents}} \\
\midrule
n-gram  & 0.753 & 0.712 & 0.688 & 0.666 & 0.642 & 0.626 \\
url     & \textbf{0.759} & \textbf{0.757} & \textbf{0.757} & \textbf{0.754} & \textbf{0.754} & \textbf{0.752} \\
title   & 0.633 & 0.630 & 0.615 & 0.588 & 0.579 & 0.558 \\
query   & 0.664 & 0.651 & 0.647 & 0.645 & 0.644 & 0.637 \\
\midrule
\multicolumn{7}{@{}l}{\emph{Newly added documents}} \\
\midrule
n-gram  & 0.753 & \textbf{0.695} & \textbf{0.677} & \textbf{0.632} & \textbf{0.627} & \textbf{0.607} \\
url     & \textbf{0.759} & 0.325 & 0.296 & 0.263 & 0.208 & 0.168 \\
title   & 0.633 & 0.362 & 0.341 & 0.336 & 0.322 & 0.308 \\
query   & 0.664 & 0.341 & 0.335 & 0.312 & 0.318 & 0.306 \\
\bottomrule
\end{tabular}
\label{tab:text_compare}
\end{table}

\subsection{Ablation study of text-based docid}
\label{sec:docid-type}

To examine the key factors of text-based docid design, we conduct an ablation study across three aspects:
(i) Docid type, where we compare different docid types, such as title, URL, n-grams, etc.
(ii) Docid granularity, where we compare docids defined at different levels of granularity, including document, paragraph, sentence, and n-gram levels.
(iii) Docid lexical diversity, where we compare the number of possible tokens used in different docid designs.

\header{Docid type}
First, we examine various forms of text-based docids, including URLs, titles, pseudo queries, and n-grams, to evaluate their effectiveness over dynamic corpora.

Table~\ref{tab:text_compare} shows the results on NQ dataset for retrieval on initial documents and new documents. We observe that the n-gram representation of docids outperforms other text-based representations with newly added documents, and also shows competitive results for initial documents in dynamic corpora.
Furthermore, this advantage stems from the inherent alignment between the n-gram representation and the pre-training objectives of language models, as both explicitly model local semantic structures through sequential token prediction, enabling n-gram docids to better adapt to the model's retrieval behavior in dynamic corpora.

\noindent\textbf{Docid granularity.}
We can define text-based docids at different levels of granularity. For example, when using the title, we essentially define the docid at the document level, where a single docid represents the entire document. In contrast, for an n-gram-based docid, any substring at any position within the document can serve as a docid to characterize the document, making it a special case of a \textit{multi-docid} design.
A finer-grained docid design may enable more nuanced matching between queries and documents and provides greater flexibility in docid generation strategies. But it also introduces a higher memorization overhead for the retrieval model.

\begin{figure}[t]
  \centering
  \includegraphics[width=0.9\linewidth]{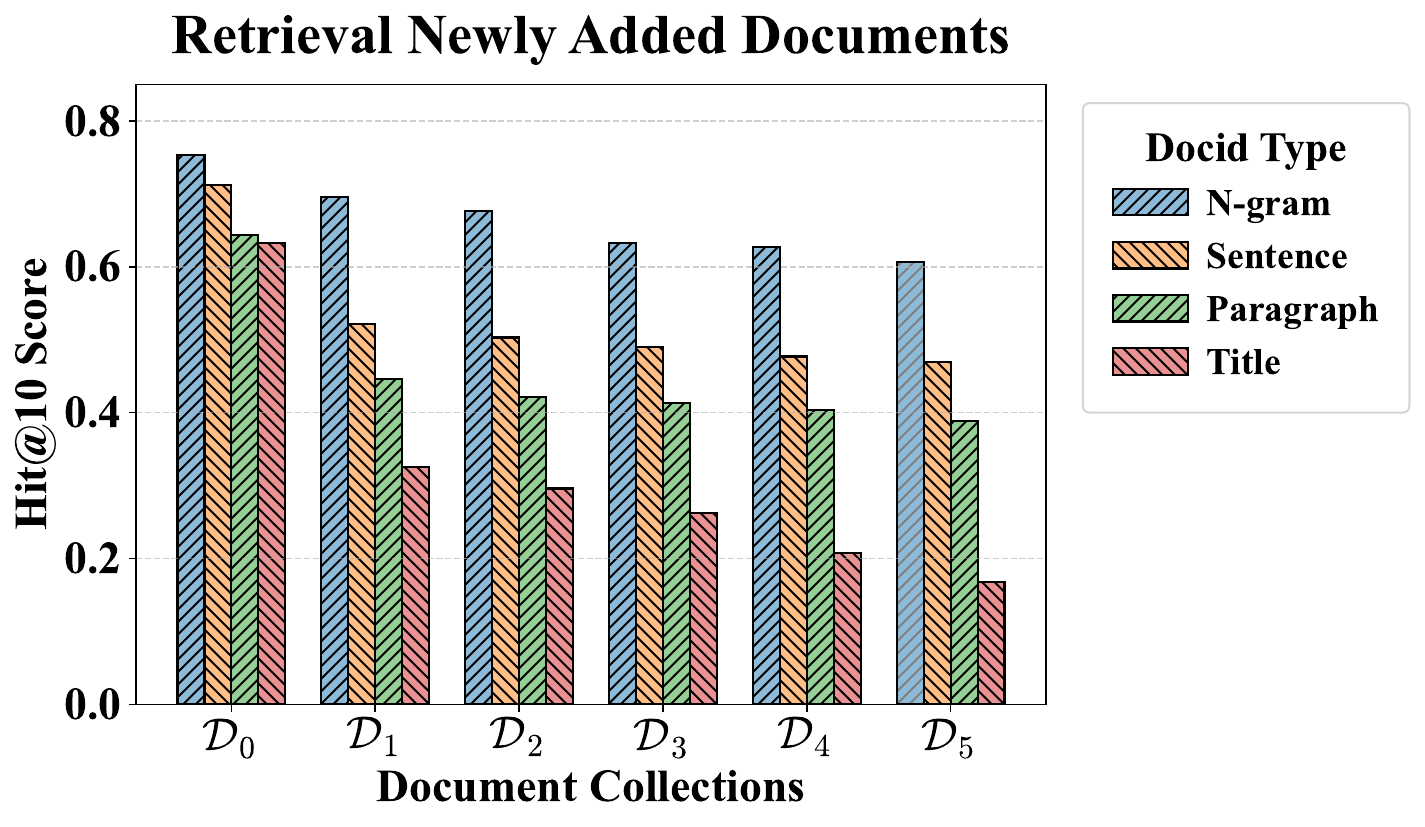}
  \caption{Hit@10 performance of n-gram docid and three fixed-position docid approaches on the NQ dataset.}
  \label{fig:position}
\end{figure}

To evaluate the impact of different docid granularities on model generalization, we define various text-based docids as follows:
\begin{enumerate*}[label=(\roman*)]
    \item \textit{Document-level}: The title is used as an abstract representation of the document.
    \item \textit{Paragraph-level}: Complete paragraphs serve as docids, and the document ranking is determined by the highest-ranked retrieved paragraph.
    \item \textit{Sentence-level}: Similar to the paragraph-level approach, but each sentence is treated as a retrieval unit.
    \item \textit{N-Gram-level}: The most fine-grained docid design, where any continuous text span within a document is used as a docid. We adopt the training and inference techniques proposed in SEAL.
\end{enumerate*}

Figure~\ref{fig:position} demonstrates that the n-gram docid approach achieves the highest Hit@10 performance, significantly outperforming the document-level (title), paragraph-level, and sentence-level baselines. This result highlights the effectiveness of the fine-grained n-gram-based multi-docid design, which enables more flexible and precise query-document matching.

\header{Docid lexical diversity}
Another important factor of text-based docids is lexical diversity.
Despite language models using a fixed-size vocabulary, different docids usually exhibit varying levels of lexical diversity.
For example, for title-based docids, the model is forced to memorize document-docid mappings rather than generalize.
When using titles as docids, the model associates specific phrases (e.g., ``Climate Change Impacts'') with fixed document identifiers. During inference, this rigid mapping fails when new documents introduce title variants (e.g., ``Global Warming Effects''). 
In contrast, the high-dimensional space of n-gram docids encourages the model to learn distributional patterns rather than discrete mappings. By exposing the model to diverse substrings during training, it becomes more robust to lexical variations in new documents.

\begin{table}[t]
\centering
\caption{Dimensional comparison n-gram vs. other types.}
\label{tab:dim}
\begin{tabular}{lcccc}
\toprule
Docid Type & Title & URL & N-gram & Numeric \\
\midrule
Size & 5,736 & 17,195 & 53,544 & 32–10,000 \\
\bottomrule
\end{tabular}
\end{table}

To measure the lexical diversity of different docids, we compute the \textit{effective vocabulary size}, i.e., the number of unique words utilized by the docids of all documents in the corpus.
As quantified in Table~\ref{tab:dim}, n-gram docids use effective vocabulary sizes that are 10–100× larger than Title or URL docids and 100–1000× larger than numeric-based docids. 
This expanded dimensional space ensures that even high-frequency n-grams in initial documents occupy only a small fraction of the total capacity, forcing the model to learn generalized substring generation patterns rather than memorizing fixed mappings.

\section{Enhancing Numeric Docids}

While text-based docids (e.g., n-grams) exhibit superior generalization in dynamic corpora, they have several limitations: (i) high computational overhead due to complex constrained decoding (e.g., FM-index traversal for n-gram matching), (ii) memory inefficiency from storing massive text fragments, and (iii) incompatibility with non-textual data (e.g., tables, images).
Conversely, numeric-based docids offer inherent advantages in storage efficiency (fixed-length numbers) and decoding speed (simple token generation), but their fixed semantic encoding limits dynamic adaptability.
This leads to the natural question of how to combine the advantages of both methods to design a docid scheme that is both efficient and effective.

Previous analyses highlight several key factors for improving GR performance in dynamic corpus scenarios:
(i) The docid should consider the \textit{semantic familiarity} to the GR model and avoid using overly unfamiliar docids.
(ii) The docid can be designed with finer granularity.
(iii) The docid can be designed with higher lexical diversity.

Based on these insights, we explore techniques to improve \textit{numeric-based} docid approaches. Our methods focus on three key aspects:
\begin{enumerate*}[label=(\roman*)]
\item We investigate how vocabulary size influences the performance of numeric-based docids and conclude that a larger vocabulary size leads to improved performance.
\item We explore a finer-grained docid design by assigning a docid to each chunk of a document rather than a single docid to the entire document.
\item We address the semantic familiarity problem by assigning known docids to new documents and introducing a novel inference strategy to enhance the model’s ability to generalize to unseen documents.
\end{enumerate*}

Combining these three aspects, we propose the \textbf{Multi-Docid Generative Retrieval (MDGR)} framework, a novel numeric-based docid design that retains the efficiency advantages of previous approaches while addressing their limitations in dynamic corpora scenarios.

\begin{figure}[t]
  \centering
  \includegraphics[width=0.85\linewidth]{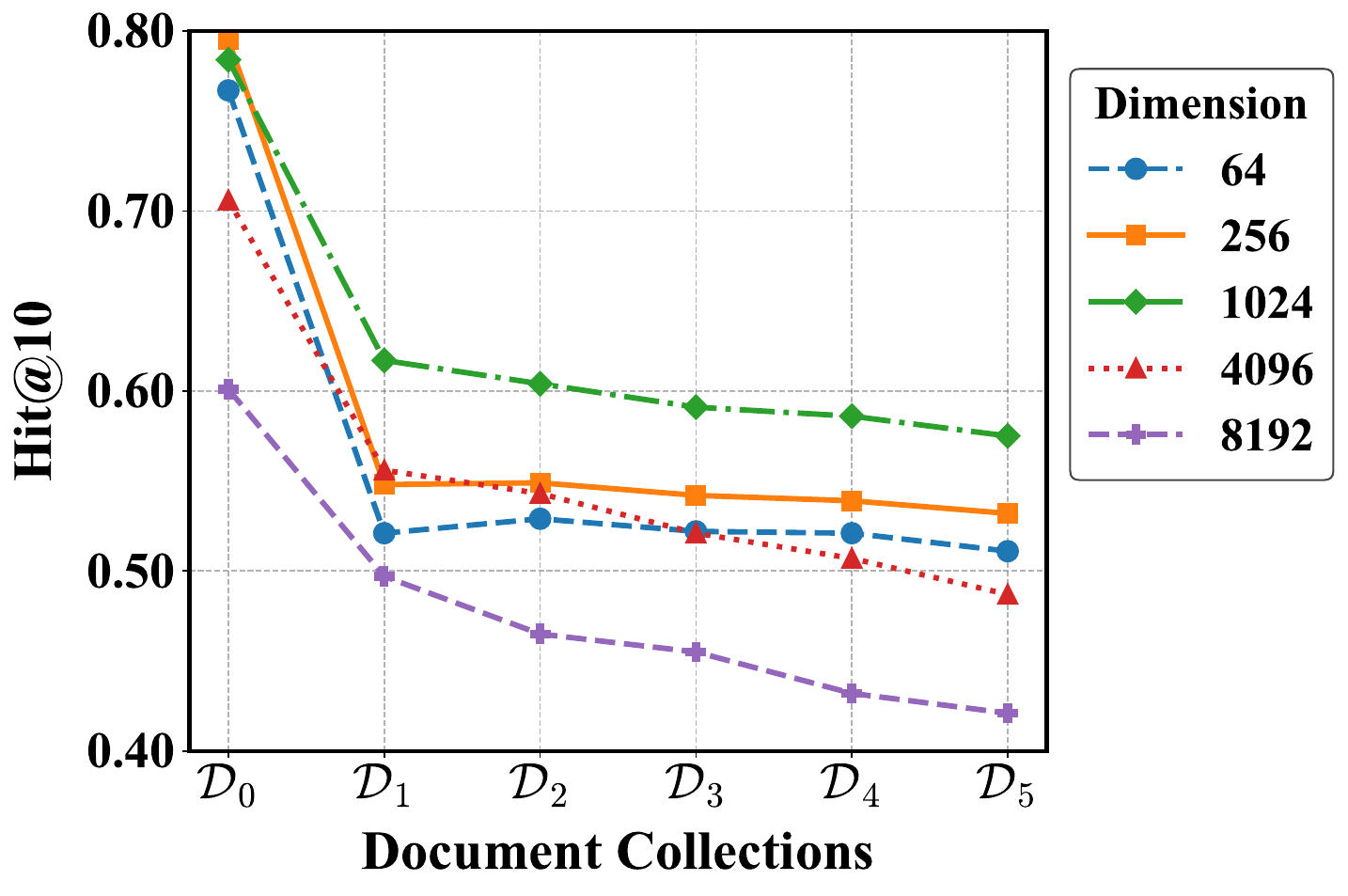}
  \caption{Hit@10 performance of different decoding dimension of Ultron-PQ on NQ dataset}
  \label{fig:dimension}
\end{figure}

\subsection{Docid design of MDGR}
\header{Optimizing vocabulary size}
Building on the insights from analysis of lexical diversity of text-based docid, we hypothesize that expanding the decoding dimension for number-type docids in GR models could similarly influence their adaptability to dynamic corpora.
To test this hypothesis, we adapt the Ultron-PQ framework by modifying its PQ parameters - the number of clusters in each subspace.
In PQ-based docid generation, document embeddings are partitioned into $m$ subvectors, each quantized into $k$ clusters via k-means.
By varying $k$ while fixing $m$, we systematically control the decoding dimension: our experiments explore $k \in \{64$, $256$, $1024$, $4096$, $8192\}$ with $m=4$.

As shown in Figure~\ref{fig:dimension}, the experimental results highlight two key patterns in docid dimension scaling. First, expanding the docid size from \(k=64\) to \(k=1024\) shows minimal impact on initial document retrieval but substantially improves performance on new documents.
Second, excessive dimensions (\(k=4096, 8192\)) cause sharp declines across all test sets. 
This could due to oversized size create sparse, under-trained docid mappings, which cause the model fails to establish reliable semantic-code relationships, particularly for new documents. 
Based on these findings, we use docid size of $k=1024$ for our following study.

\header{Designing fine-grained docids}
We explore a multi-docid approach to designing numeric-based docids with finer granularity.  
Specifically, we partition a document \( d \) into semantic chunks \( \{c_1, c_2, \ldots, c_N\} \) using a sliding window approach, where each \( c_i \) represents a text fragment of the document.  
Each chunk is independently mapped to a numeric-based docid \( z_i \) through product quantization (PQ).  
This results in multiple number sequences per document, analogous to the multi-docid design of n-grams.

\header{Constrained docid expansion}
To address the semantic gap in numeric-based docids, we propose a constrained docids expansion strategy on dynamic corpora. 
We store all docids from the initial document collection, and constrain the indexer to use only these existing docids when indexing new document chunks.

\subsection{Inference strategy of MDGR}
To retrieve the relevant documents, we propose a simple document ranking strategy for our approach. 
Given a query, we first perform constrained beam search to generate several candidate docids $ \{z_1, z_2, \dots, z_k\} $. We then retrieve all documents containing at least one of these docids. For each candidate document $ d_j $, we compute a score
based on the weighted sum of the beam search positions of the contained docids. The score for document $ d_j $ is given by:
\begin{equation}
\text{Score}(d_j) = \text{Coverage Count}(d_j) + \beta \times \sum_{z_i \in d_j} \frac{1}{\text{rank}(z_i)},
\end{equation}
where $\text{Coverage Count}(d_j) $ is the number of distinct docids in $ d_j $; $ \text{rank}(z_i) $ is the position of docid $ z_i $ in the beam search (with 1 being the highest rank); and $\beta $ is a hyperparameter that controls the importance of the ranking term relative to the coverage count.

\subsection{Implementation and evaluation results}
\begin{table}[t]
\centering
\setlength{\tabcolsep}{1.2mm}
\caption{Comparison of retrieval performance (Hit@10) on
the NQ dataset for newly added documents.}
\label{tab:methods}
\begin{tabular}{l cccccc}
\toprule
\multirow{2}{*}{Method} & \multicolumn{6}{c}{NQ (Hit@10)} \\
\cmidrule(lr){2-7} & $\mc{D}_0$    & $\mc{D}_1$    & $\mc{D}_2$    & $\mc{D}_3$    & $\mc{D}_4$    & $\mc{D}_5$ \\ 
\midrule
BM25      & 0.647 & 0.620 & 0.588 & 0.598 & 0.552 & 0.571 \\
DPR-HN    & 0.826 & 0.645 & 0.644 & 0.626 & 0.621 & 0.624 \\
NCI       & \textbf{0.871} & 0.464 & 0.437 & 0.433 & 0.358 & 0.323 \\
SEAL      & 0.809 & \textbf{0.744} & \textbf{0.736} & \textbf{0.727} & \textbf{0.727} & \textbf{0.725} \\
MDGR      & 0.824 & 0.717 & 0.704 & 0.695 & 0.647 & 0.633 \\
\midrule
\multicolumn{7}{l}{Ablation Study} \\
w/o constrain   & 0.824 & 0.447 & 0.424 & 0.408 & 0.377 & 0.356 \\
w/o multi-docid  & 0.831 &	0.523 &	0.511 &	0.497 &	0.488 &	0.473 \\
MDGR (size=64)  & 0.843 &	0.472 &	0.436 &	0.417 &	0.409 &	0.381 \\
MDGR (size=8192) & 0.674	& 0.574 &	0.565 &	0.533 &	0.531 &	0.512 \\
\bottomrule
\end{tabular}
\end{table}

Our model is implemented using the T5-base architecture.  
We initialize the model with pre-trained weights from Hugging Face's T5-base checkpoint.  
Following previous work, the training objective includes three parts:  
\begin{enumerate*}[label=(\roman*)]
    \item \textbf{Training with synthetic queries.} Minimize the loss between generated pseudo-queries and their corresponding docids.  
    \item \textbf{Encoding document contexts.} Minimize the loss between each document chunk \( c_i \) and its original docid.  
    \item \textbf{Incorporating real queries.} Minimize the loss between user queries and target docids.  
\end{enumerate*}

For training, we employ the AdamW optimizer with a base learning rate of $1 \times 10^{-5}$ and a linear warmup over the first 10\% of training steps. We set the batch size to 64 and train for 10 epochs, which requires approximately 4 hours on 8 NVIDIA A100 GPUs. 
To ensure efficient document processing, we split documents into chunks using a sliding window approach: each chunk contains 256 tokens with a stride of 128 tokens.
To get the numeric-based docids, we first generate semantic embeddings for text chunks using a frozen BERT-base model, then apply product quantization to convert these continuous vectors into docids.
The docids have a size of 1024 and a length of 4.

As shown in Table~\ref{tab:methods}, the MDGR framework exhibits competitive effectiveness in comparison to existing retrieval models.
While the performance of MDGR is not the absolute best across all document sets, it achieves solid results over dynamic corpora.

We also show some ablation variant to evaluate three critical design of our method:
First, removing the constrained docid expansion strategy leads to significant deterioration, demonstrating that generating new docids for updated documents would disrupt the semantic consistency of numbering-based docids.
Second, when disabling multi-docid indexing (single-docid per document), the performance drops significantly, confirming our hypothesis that fine-grained docid design can better preserves granular semantic associations.
Furthermore, we investigate the impact of docid size.
Small docid sizes (64) lead to substantially degrade on newly added documents.
Conversely, excessively large sizes (8192) create an over-discretized learning space where the model struggles to establish stable query-docid mappings.
\begin{table}[t]
\centering
\caption{Experiments about memory costs and efficiency.}
\renewcommand{\arraystretch}{1}
\setlength\tabcolsep{12pt}
\label{tab:memory}
\begin{tabular}{lccc}
\toprule
\textbf{Method} & \textbf{Memory} & \textbf{Tok-K} & \textbf{Latency} \\
\midrule
DPR & \phantom{0}980MB & 100 & 152ms \\
\midrule

\multirow{2}{*}{NCI} & \multirow{2}{*}{\phantom{0}865MB} & \phantom{0}10 & 216ms \\ 
& & 100 & 269ms \\ 
\midrule

\multirow{2}{*}{SEAL} & \multirow{2}{*}{2200MB} & \phantom{0}10 & 619ms \\ 
& & 100 & 778ms \\ 
\midrule

\multirow{2}{*}{MDGR} & \multirow{2}{*}{\phantom{0}886MB} & \phantom{0}10 & 241ms \\ 
& & 100 & 320ms \\ 
\bottomrule
\end{tabular}
\end{table}

Table~\ref{tab:memory} shows MDGR achieves both effectiveness and system efficiency. This efficiency stems from combining number-type docids' compact storage with number sequences, enabling dynamic adaptability without sacrificing speed – MDGR retrieves documents 3.6× faster than SEAL while outperforming other numeric-based docid methods (NCI) in dynamic corpora scenario.

\section{Conclusion}
In this paper, we have conducted a systematic investigation into the capabilities and limitations of GR models in dynamic corpora scenarios, where document collections expand continuously over time. Through comprehensive evaluations on realistic dynamic benchmarks derived from NQ and MS-MARCO datasets, we have revealed that certain GR methods, primarily those relying on numeric-based docids, face challenges in generalizing to unseen documents without additional training, while models utilizing text-based docids demonstrate stronger generalization capabilities in dynamic corpora.
Our analysis has demonstrated that text-based docids inherently address key limitations of numeric-based docids by mitigating generation bias towards previously seen docids, enabling finer-grained document representation, and enhancing robustness against overfitting.

Building on these insights, we have proposed a novel GR framework MDGR that combines the structured benefits of numeric-based docids with a revised design to avoid training-induced biases, achieving competitive performance in dynamic corpora scenario. Our findings not only highlight the critical role of docid semantics and representation in GR frameworks but also provide actionable guidelines for adapting generative retrieval to real-world dynamic environments. Future work may explore hybrid docid strategies or pretraining-enhanced generalization mechanisms to further bridge the gap between static and dynamic retrieval performance.
\begin{acks}

This research was (partially) funded by
    the Natural Science Foundation of China (62472261),
    the Provincial Key R\&D Program of Shandong Province with grant No. 2024CXGC010108, 
    the Technology Innovation Guidance Program of Shandong Province with grant No. YDZX2024088, 
    the Dutch Research Council (NWO), under project numbers 024.004.022, NWA.1389.20.\-183, and KICH3.LTP.20.006, 
    and 
    the European Union's Horizon Europe program under grant agreement No 101070212.   
All content represents the opinion of the authors, which is not necessarily shared or endorsed by their respective employers and/or sponsors.

\end{acks}

\bibliographystyle{ACM-Reference-Format}
\balance
\bibliography{references}

\end{document}